\documentclass[superscriptaddress,aps,twocolumn,amsfonts,nofootinbib]{revtex4-1}
\begin{document}

\title{Ghost condensation and the Ostrogradskian instability on low derivative backgrounds}
\author{Justo L\'opez-Sarri\'on}
\email{jujlopezsa@unal.edu.co}
\affiliation{Departamento de F\'{\i}sica Te\'orica, At\'omica y \'Optica, Universidad de Valladolid,
47011 Valladolid, Spain}

\affiliation{Departamento de F\'isica, Universidad Nacional de Colombia\\
Ciudad Universitaria, K. 45 No. 26-85, Bogot\'a D.C., Colombia}

\author{Mauricio Valencia-Villegas}
\email{j.valencia@physik.uni-muenchen.de,\\
jmvalenciavillegas@gmail.com}
\affiliation{Arnold Sommerfeld Center for Theoretical Physics,\\ Ludwig-Maximilians-Universit\"at M\"unchen, Theresienstrasse 37, 80333, Germany}

\begin{abstract}
We show a new class of interaction terms with higher derivatives that can be added to every low derivative real scalar, such that the theory is degenerate, and the equation of motion remains of second order. In contrast to previous setups, the necessary constraints that eliminate the ghosts  also  have a clear physical motivation: they impose and preserve at all times the low derivative dynamics of the real scalar as a background, on top of which, the fluctuations induced by these higher derivative terms are degenerate, and ghost-free. We summarize the setup for these constrained ghost-free real scalars in a two-step prescription. In contrast to some models with first-order derivative interactions  with applications for dark energy and inflation, these theories with  necessarily constrained second-order derivative self-interactions do not modify the speed of propagation, neither the cone of influence for the field equations, which are prescribed by the low derivative background that is also essential to eliminate the ghost; hence, avoiding the potential superluminality issues of the former.
\end{abstract}

\maketitle
\section{Introduction}

In spite of the phenomenological success of the standard model of particle physics (SM) and Einstein gravity in a wide range of energies, it is understood that they do not give the full picture. New effects are explored beyond the currently tested energy scales, for instance, by adding effective terms to the SM which must  cause  negligible corrections in at least some energy regimes. In striking contrast, non-degenerate higher time derivative terms do not induce small corrections but  radically modify the physics through non-perturbative effects, as they bear a fundamental instability \cite{simon}. They enlarge the dimensionality of phase space including a ghost that catasthropically destabilizes the low derivative degrees of freedom ({\it dof}'s) upon interaction \cite{1woodard,2woodard,3woodard,noui,simon,tolley}. Hence, higher derivative extensions  are either neglected or only applied in the regime of the effective theory, for instance, by the method of ``perturbative constraints'', which can at best hide the ghost at some order, but not eliminate it \cite{1woodard,2woodard,molina,3woodard,trodden,taiwan,glavan,simon,carsten,burgess}.

Galileons \cite{galileons}, and degenerate theories \cite{tolley,1langlois,klein,1derham,2langlois} are among some exceptions \cite{lovelock,horndeski,deffayet,3langlois}. The theorem of Ostrogradsky states that every  non-degenerate higher derivative theory entails an unbounded energy from below, which is ultimately seen in the propagation of an additional ghosty  \textit{dof} \cite{ostrogradsky, 1woodard,noui,simon}. A comprehensive introduction to the Ostrogradsky's instability is given, for instance, in \cite{1woodard}, or in \cite{1motohashi}, where further results on this instability with multiple higher derivative dynamical variables are shown.

For the case of interest in this letter with only one higher derivative dynamical variable, the constrained degenerate theories circumvent the issue at the expense of introducing an additional {\it ad-hoc} dynamical variable with low derivatives, and by imposing constraints by fiat among the latter  that eliminate the ghosty {\it dof} \cite{1langlois}. In brief, a single higher derivative dynamical variable usually lacks physical interpretation of its own, because in general, it needs of an additional auxiliary low derivative variable.

In this letter, we show a new class of interaction terms with higher derivatives that can be added to every low derivative real scalar, such that the theory for this sole higher derivative variable is degenerate, and the nonlinear equation of motion remains of second order. In other words, we give a restriction on the type of higher derivative terms that do not carry ghosts even at non-perturbative level, once a physically motivated constraint is imposed. Hence, they may be more suitable to explore higher order terms in effective theories, as there is no need to argue that ghosts are hidden beyond the regime of validity of the effective theory \cite{molina,trodden,taiwan,glavan,simon,carsten,burgess}.

As it is standard in degenerate theories, we introduce a constraint, and an auxiliary variable. However, for this novel class, the latter two ingredients are not merely a technicality to eliminate ghosts, and instead they have a physical meaning deeply related to the sole higher derivative variable. Namely, the necessary constraints also  have a clear physical motivation: they impose and preserve at all times the low derivative dynamics of the scalar as a background, on top of which, the fluctuations induced by these higher derivative terms are degenerate, and ghost-free. On the other hand, the auxiliary quantity is expressed on-shell as a function of the single higher derivative variable.

Finally, the dimensionality of phase space is at most the same as for the corresponding low derivative theory, in accordance with  \cite{1langlois,tolley}. In other words, there are redundancies, which, once removed, would let us see that the dynamics is fundamentally equivalent to a  conventional low derivative theory. However, if the low derivative equivalent theory is explicitly written, it takes in principle a prohibitively difficult form, and physical significance may be obscured.\\

We proceed as follows: in section \ref{sec assumptions} we state the assumptions and notation. In section \ref{sec class} we motivate the reduced class of degenerate theories where the two ingredients, namely, the auxiliary variable, and the constraint have the physical meaning discussed above. Then, in sections \ref{sec new d} and \ref{sec first stability},  respectively, we briefly summarize the implications for the  dynamics, and we give a first analysis of the energy function for this setup. In section \ref{sec general} we summarize the construction of these degenerate ghost-free theories in a two-step prescription.

The Ostrogradsky's stability is verified in section  \ref{sec stability}. This section clarifies the elimination of ghosts. The strategy is as follows: First, the Hamiltonian is written following Ostrogradsky's construction \cite{1woodard,ostrogradsky}. Then, because there are constraints, we must follow Dirac's programme. That is: we fix the low derivative sector of the dynamical variable via constraints despite the higher derivative terms. This must be satisfied at all times if it is satisfied at an initial time. In general, this concistency requirement (time conservation of the constraint) is independent of the constraint itself, and hence, it implies further constraints in the dynamics. In fact, the latter must also be preserved at all times. This programme reveals the complete constrained structure of the theory, such that all constraints are satisfied at all times. A relevant implication is that the dimensionality of phase space is reduced, and there are less degrees of freedom than it is apparent from the number of dynamical variables and higher derivative terms in the Lagrangian. Now, for the specific class of theories summarized in section \ref{sec general} we will count the ghost among the degrees of freedom that are eliminated.

In brief, from the {\it first constraint} that fixes the low derivative dynamics of the variable as a background at an initial time, all consistency constraints that hold at all times are deduced, and it is the latter that eliminate the ghost. 

Note that the ghost-elimination hinges from fixing in first place a {\it low derivative background}. This is such that there can be an energy regime where the background -low derivative mode- is more relevant for the dynamics than the ghost-free fluctuations caused by the higher derivative terms. A  primary motivation for such a first constraint is the phenomenological success of low derivative theories at least at low energies. 

In section \ref{sec low}, we show the low dimensionality of phase space, and most importantly, we show that {\it the dynamics for this restricted class of theories fundamentally does not include higher derivatives}, which can be explicitly seen when the theory is written in terms of a different set of canonical coordinates. On this regard, at the end of section \ref{sec low} we argue on the possible usefulness of this construction without resorting to the set of variables in which the theory is manifestly low in derivatives.\\

On the other hand, the necessary auxiliary variable in this setup is simply a Lagrange multiplier imposing the physically motivated constraint that was discussed above. Such case, where the auxiliary variable required to eliminate the ghost is a Lagrange multiplier, was already explored in \cite{tolley}, although it was restricted to nonlinear terms that are at most quadratic in the second time derivative of the dynamical variable. The cases shown in this letter are a non-exhaustive extension to the latter. 

For instance, in section \ref{sec example} we show a simple concrete example for the mechanics of a single particle where this auxiliary  Lagrange multiplier is expressed as the superposition of the low derivative modes, and the fluctuations due to higher derivative terms, suppressed by a new scale. This auxiliary variable carries the higher derivative effects solving a forced,  second order differential equation (no ghosts), hence, it is called the {\it ghost condensate}. 

We compute the solution for the ghost condensate in this example. We  show that the fluctuations about the low derivative background are suppressed by the new-physics scale at which the higher derivative terms would become relevant. Thus, when the new scale is arbitrarily large, we recover the initial low derivative theory, and the ghost condensate becomes the usual background solution. Finally, we also show how the low dimensionality of phase space is intimately related to the elimination of the ghosty degree of freedom, and that it is associated to the boundedness from below of the energy function in this example.\\ 

Finally, in section \ref{sec field} we specialize to real scalar fields. In particular, let us bring to attention some of the issues that low derivative self-interacting scalars may have \cite{2arkani-hamed,mukhanov,susskind,shore,visser,2derham,ellis}: for first-order derivative interactions there are nonperturbative effects that can cause the breakdown of the Cauchy problem, loss  hyperbolicity \cite{susskind}, as well as causality issues, because there is a modified (acoustic) cone of influence \cite{ellis}. These issues may, for instance, be problematic in k-essence models \cite{mukhanov}. Then, the main purpose in section \ref{sec field} is to stress on the potentially benign properties of these {\it necessarily constrained} higher derivative self-interactions on these regards. We stress on the speed of propagation, the (acoustic) cone of influence for the wave equation, and subluminality despite the built-in higher derivative self-interactions, as opposed to low derivative {\it unconstrained} interactions.

In section \ref{sec discussion} we discuss a possible interpretation of the constraint, and the auxiliary variable in this setup. We give the conclusions in section \ref{sec conclusions}.

\section{Fluctuations about low derivative dynamics}\label{sec small fluctuations} 
The key assumption that permits the Ostrogradskian instability in the dynamics is {\it nondegeneracy} \cite{1woodard,2woodard}. In this letter we follow the strategy of introducing constraints and an auxiliary variable with low derivatives, in order to violate this assumption. In other words, we make the higher derivative theory {\it degenerate}, so to avoid the Ostrogradskian instability \cite{1langlois}.

In general, the two ingredients, namely, the constraint, and the auxiliary variable are a mere  technical necessity. In this section, we intend to give a physical motivation for a specific class of constraints, and to identify the allowed class of higher derivative terms in the effective theory. It will facilitate the physical interpretation of the auxiliary variable, and of the corresponding degenerate theories in section \ref{sec discussion}.

As it was discussed in the introduction, nondegenerate higher time derivative terms do not induce small corrections in an effective theory but  {\it radically modify the low derivative theory} through non-perturbative effects \cite{simon}. Therefore, the strategy in section \ref{sec class} will be to consider fluctuations about solutions to the low derivative theory (the fluctuations caused by the addition of higher derivative terms). In general, these fluctuations solve higher than second order equations.  Thus, among the possible higher derivative terms that were added, we will identify those with the property that the  fluctuations are degenerate on this specific low derivative background.

Then, the background will be fixed in the full dynamics via a constraint and its time conservation. We will go on to verify in section \ref{sec stability} that once we add these specific higher derivative terms to the initially low derivative theory, then the complete theory with  the above mentioned constraint is degenerate. As expected, in section \ref{sec low}, we verify the equivalence of this {\it constrained} higher derivative theory to a low derivative theory written in terms of another set of canonical coordinates.

All in all, the constraint not only eliminates ghosts for the specific class of higher order terms to be shown below, but also has a ``natural'' phenomenological motivation: to hold the solutions to the low derivative theory as a relevant contribution to the dynamics despite the higher derivative terms.   This is further discussed in section \ref{sec discussion}.

The notation and assumptions are given in section \ref{sec assumptions}.

\subsection{The assumptions: a basis of unstable higher derivative theories}\label{sec assumptions}
Let us first consider a theory with Lagrangian depending on up to first time derivatives of the dynamical variable  $\phi$, and possibly, spatial derivatives $\mathcal{L}^{(1)}(\partial \phi,\phi)$. For the moment, unless stated otherwise in the next sections, and to avoid distractions with non-essential field formalism, we will first analyze mechanics of the sole dynamical variable $\phi$ depending only on time. 

We will assume that {\it only}\footnote{Let us stress that demanding nondegeneracy for the low derivative sector $\mathcal{L}^{(1)}$ poses no condition on the high derivative sector. In particular, we will impose degeneracy on the high derivative sector by means of constraints.} this low derivative sector $\left(\mathcal{L}^{(1)}\right)$ is nondegenerate. That is, with $\dot{\phi}$ time derivative of $\phi$,
\begin{eqnarray}
\frac{\partial^2 \mathcal{L}^{(1)}}{\partial \dot{\phi}^2} \neq 0 \, , \label{eqn non degeneracy L1}
\end{eqnarray}
such that the Euler-Lagrange equation for $\mathcal{L}^{(1)}$ depends linearly on $\ddot{\phi}$. Since two initial conditions must be given in order to solve the second order equation of motion, the dimensionality of phase space is two. Thus, we are implicitly assuming that with $\mathcal{L}^{(1)}$ we describe only one \textit{dof}.

Now, adding higher derivative terms to the low derivative theory $\mathcal{L}^{(1)}$,
\begin{eqnarray}
\mathcal{L}=\mathcal{L}^{(1)} + \mathcal{L}^{(2)} \, , \label{eqn L without constraint}
\end{eqnarray}
where  $\mathcal{L}^{(2)}(\partial^2 \phi,\partial \phi,\phi)$ depends  on second time derivatives of $\phi$, increases  the number of  {\it dof}'s (introduces a ghost) in the case that $\mathcal{L}^{(2)}$ is nondegenerate: namely, the higher derivative theory $\mathcal{L}$ is nondegenerate if the term
\begin{eqnarray}
\frac{\partial^2 \mathcal{L}^{(2)}}{\partial \ddot{\phi}^2} \,,\label{eqn non degeneracy L2}
\end{eqnarray}
does not vanish. With this non degenerate condition, the theory (\ref{eqn L without constraint}) is unstable and non unitary upon quantization \cite{1woodard,2woodard,3woodard,noui,simon, tolley,1langlois,klein,2langlois}. For definiteness, we will assume: (\ref{eqn non degeneracy L2}) is not {\it identically} zero, but, in order to satisfy the degeneracy condition, we will construct a constrained version of (\ref{eqn L without constraint}) where (\ref{eqn non degeneracy L2}) {\it vanishes only on-shell}, on the surface of constraints in phase space. 

Furthermore, for the moment, we consider $\mathcal{L}^{(2)}$ strictly high on derivatives. That is, $\mathcal{L}^{(2)}$ can be written as a function proportional to $\ddot{\phi}$, and these higher derivatives cannot be eliminated by integration by parts.

Finally, we will canonically normalize $\phi$ taking as reference the standard low derivative kinetic term in $\mathcal{L}^{(1)}$, in such a way that with $\mathcal{L}^{(2)}$, we introduce a new  scale ($\Lambda$). Two concrete examples are given in section \ref{sec example}, in equations (\ref{eqn L1 example}), (\ref{eqn L2 example}) and (\ref{eqn Lprime example}), and in section \ref{sec field}, for a real scalar field in four dimensions, where $\Lambda$ is a new energy scale. 

Note that, if the higher derivative terms $\mathcal{L}^{(2)}$  are not degenerate, they  cannot induce small corrections to the low derivative dynamics of $\mathcal{L}^{(1)}$. First, the equation of motion is of 4th order and four initial conditions are required; hence, the dimension of phase space for dynamics of the theory (\ref{eqn L without constraint}) is larger than in the theory $\mathcal{L}^{(1)}$. Second, upon interaction, the low derivative \textit{dof} that would be propagated with $\mathcal{L}^{(1)}$ dynamics  is catastrophically  destabilized. In other words, (\ref{eqn L without constraint}) leads to unstable dynamics regardless if there is any small parameter ({\it e.g.} $\Lambda^{-1}$) suppressing the higher derivative sector $\mathcal{L}^{(2)}$ \cite{simon}. As noted long ago, the Ostrogradsky's instability is a non-perturbative effect \cite{1woodard,2woodard,3woodard,noui,simon}. Perturbative expansions for (\ref{eqn L without constraint}) can at best hide the ghost \cite{1woodard,2woodard,molina,3woodard,trodden,taiwan,glavan,simon,carsten,burgess}.

\subsection{The physically motivated constraint, and the restricted class of effective theories}\label{sec class}
Below, we construct a class of \textit{constrained}    modifications of the {\it unstable theory} (\ref{eqn L without constraint}), which we denote as $\mathcal{L}'(\phi)$. They have the property that fluctuations about the background set by the low derivative dynamics are degenerate, stable in the sense of Ostrogradsky, and can be arbitrarily small. In other words,  the higher derivative terms in the constrained theory $\mathcal{L}'(\phi)$,
\begin{enumerate}
\item induce only corrections to the low derivative modes described by $\mathcal{L}^{(1)}$ (no new degrees of freedom are integrated-in). These corrections are suppressed by $\Lambda$, and,
\item  they are naturally stabilized by the low derivative modes described by $\mathcal{L}^{(1)}$ (the fluctuations induced by $\mathcal{L}^{(2)}$ are degenerate on the $\mathcal{L}^{(1)}$-background).
\end{enumerate}
Note that we are selecting $\mathcal{L}^{(1)}$ as special for the dynamics. A primary motivation is the phenomenological success of low derivative theories at least at low energies. Hence, we will  refer to the low derivative dynamics derived from $\mathcal{L}^{(1)}$ as the {\it low derivative sector}, as well as the {\it low energy sector}, and the corresponding solutions as the {\it low derivative modes}, or {\it low energy modes}.\\

By the properties of the corrected theory $\mathcal{L}'$, we consider a perturbative expansion with the distinctive feature that the leading, 0-th order approximation is the standard, low derivative dynamics. Namely, denoting the Euler-Lagrange equation for the dynamical variable $\psi$ derived from an action with Lagrangian possibly depending on more variables $F(\psi,\xi,\,\dots)$, as,
\begin{eqnarray}
\Theta(F;\psi)=0 \, , \label{eqn notation E-L}
\end{eqnarray}
we denote the solution for the {\it low derivative sector} $\mathcal{L}^{(1)}(\phi)$ as $\phi_0$  such that,
\begin{eqnarray}
\Theta(\mathcal{L}^{(1)};\phi_0)= 0 \, .\label{eqn def background}
\end{eqnarray}
That is, $\phi_0$ represents the {\it low derivative mode}. Now, let us  decompose $\phi$ in the complete theory $\mathcal{L}'(\phi)$  in terms of a fluctuation ($\pi$) about this 0-th order solution as,
 \begin{equation}
 \phi= \phi_{0}+\epsilon\pi \, ,
 \label{decomposition}
 \end{equation}
where $\epsilon$ is an \textit{arbitrarily small} dimensionless parameter, whose physical meaning will be argued below. Note that this expansion is meaningless for the unstable theory (\ref{eqn L without constraint}). However, for the corrected theory $\mathcal{L}'$, the expansion must be meaningful in the sense that small fluctuations
$$\epsilon \|\pi\|\ll\|\phi_{0}\|  $$
about the low derivative solution $\phi_0$ must not become large arbitrarily fast, as opposed to an Ostrogradsky unstable theory (\ref{eqn L without constraint}).

Let us first expand about $\phi_0$ in (\ref{eqn L without constraint})  up to  order $\mathcal{O}(\epsilon^2)$ and read out  from the signatures of the instability  the structure that $\mathcal{L}'$ must have. $\mathcal{L}$ becomes,
\begin{eqnarray}
\mathcal{L}&=&\mathcal{L}_0+\mathcal{L}_\pi \,, \label{lagrangianExpand}
\end{eqnarray}
where $\mathcal{L}_0$ can be written as a total time derivative, as it only depends on the fixed background $\phi_0$, and  $\mathcal{L}_\pi$  is,
\begin{eqnarray}
\mathcal{L}_\pi&=& \epsilon^2\frac{1}{2}\left.\frac{\partial^2\mathcal{L}^{(2)}}{\partial \ddot{\phi}^2}\right\vert_{0}\ddot{\pi}^2+\tilde{\mathcal{L}}_\pi+ \mathcal{O}(\epsilon^3)\,, \label{eqn quadratic L}
\end{eqnarray}
where the term $\tilde{\mathcal{L}}_\pi $ is, up to a total time derivative, a second order polynomial of $\epsilon\pi$ and $\epsilon\dot\pi$, and the subindex $\vert_0$ means evaluation at the background $\phi_0$.  Notice again that the $\mathcal{O}(\epsilon^2) $ approximation in (\ref{eqn quadratic L}) cannot be justified for the naive, unstable theory (\ref{eqn L without constraint}), however, it should be meaningful for the corrected theory. 

Let us take a first step to construct $\mathcal{L}'$:
 at order $\mathcal{O}(\epsilon^2) $ the only term that signals the Ostrogradsky's instability is the first on the right hand side of equation (\ref{eqn quadratic L}), namely, the term proportional to $ \ddot{\pi}^2$. It leads to a linearized equation  of 4-th order for $\pi$. Clearly, the constrained dynamics $\mathcal{L}'$ should not have such terms in the perturbative expansion about the low energy mode $\phi_0$. Let us see the options: on the one hand, if (\ref{eqn L without constraint}) contains no self-interactions in the higher derivative sector, then (\ref{eqn non degeneracy L2}) should be identically zero in order to eliminate the signature of the instability in (\ref{eqn quadratic L}). This is the trivial degeneracy requirement that we excluded in first place. 

On the other hand, if there are nonlinear terms in the higher derivative sector, then (\ref{eqn non degeneracy L2}) is a function of $\phi$ and derivatives. Since this function must be such that, if it is evaluated on the 0-th order solution $\phi_0$, it vanishes; then, a clear choice is,
\begin{eqnarray}
\frac{\partial^2\mathcal{L}^{(2)}}{\partial \ddot{\phi}^2}\propto \Theta(\mathcal{L}^{(1)};\phi) \, , \label{eqn construction L2}
\end{eqnarray}
such that the perturbative expansion for $\mathcal{L}'$ reduces to,
\begin{eqnarray}
\mathcal{L}_\pi&=&\tilde{\mathcal{L}}_\pi+ \mathcal{O}(\epsilon^3)\,, \label{eqn quadratic L'}
\end{eqnarray}
which contrasts to (\ref{eqn quadratic L}) because there are no $\propto \ddot{\pi}^2$ contributions, such that in this case the fluctuations ($\pi$) at $\mathcal{O}(\epsilon^2)$ solve a second order equation with (space-)time dependent coefficents fixed by $\phi_0$. 

Let us note that (\ref{eqn construction L2}) is not identically zero, and yet, it vanishes for fluctuations about low derivative modes at $\mathcal{O}(\epsilon^2) $. Below, we must implement this non-trivial degeneracy at order $\mathcal{O}(\epsilon^3) $. Let us stress that we must capture the nonperturbative effects of the derivative self-interactions in $\mathcal{L}^{(2)}$; hence, we will not arbitrarily split ``free'' from ``interaction'' part in the Lagrangian. This was clear in the expansion performed above, because the term (\ref{eqn construction L2}) proportional to the quadratic term $\propto \ddot{\pi}^2$ in (\ref{eqn quadratic L}) includes all contributions from both quadratic (if any), and interaction terms in $\mathcal{L}^{(2)}$. Indeed, there are well known, analogous examples where  nonperturbative effects of derivative self-interactions are key to  control, or even endanger the stability. For instance, for first order derivative self-interactions: ghost condensation  \cite{1arkani-hamed}, potential superluminality issues in k-essence \cite{mukhanov}, and stability of the Cauchy problem \cite{susskind,mukhanov}.

Furthermore, note that the particular choice in (\ref{eqn construction L2}) is not a necessary condition; however, it encloses the interesting property that the stability of the higher derivative theory hinges on the  propagation of the well-known low energy modes. Thus, one could expect that  corrections due to $\mathcal{L}^{(2)}$ could be, for instance, masked in low energy scatterings in the observation of the low derivative degrees of freedom described with $\mathcal{L}^{(1)}$, which here are taken as the $\mathcal{L}^{(2)} $-stabilizing background (so far at $\mathcal{O}(\epsilon^2)$). 

Now, Ostrogradsky's instability is a nonperturbative effect which cannot be hidden at $\mathcal{O}(\epsilon^3)$. Therefore, as it is well known, $\mathcal{L}'$ must have constraints and be degenerate \cite{tolley,1langlois,klein,1derham,2langlois,1woodard,2woodard}. Let us show the constraint in $\mathcal{L}'$ that not only stabilizes at  nonperturbative level, but also makes meaningful the above defined expansion about the degrees of freedom described with $\mathcal{L}^{(1)}$: in the unstable theory (\ref{eqn L without constraint}) the origin of the issue is that the low derivative mode $\phi_0$ is not relevant as a 0-th order solution in a perturbative expansion. In other words, fluctuations, ``small'' with respect to $\phi_0$, would become rapidly ``large'' due to the Ostrogradsky unstable $\mathcal{O}(\epsilon^3)$ terms. This suggests that  for the corrected theory $\mathcal{L}'$, it must be guaranteed in first place that the low derivative mode $\phi_0$ is a background on top of which fluctuations can be built despite the higher derivative terms. We will implement this below via constraints, and their time conservation. For instance, we show in a concrete simple example in section \ref{sec example} that fixing via constraints the low derivative mode $\phi_0$ as a background, $\pi$ can be identified to all orders as a particular solution to a {\it second order} non homogeneous equation (\ref{eqn pi}).

Let us first specify the constraint and the final form of the modified theory $\mathcal{L}'$. Then, we show in section \ref{sec stability} that this is enough to eliminate the ghost at non-perturbative level. Also, a primary analysis of the signatures of the stability is given below, in subsection \ref{sec first stability}.

On the other hand, in section \ref{sec low} we show the equivalence\footnote{However, as we will note, the equivalent low derivative theory is written in terms of a different set of canonical coordinates which may hold no obvious relation to $\phi$, neither to the physically meaningful solutions to the low derivative theory ($\phi_0$).} of $\mathcal{L}'$ dynamics to a conventional low derivative theory.\\

Altogether, in $\mathcal{L}'$ this can be treated as a genuine constraint on the dynamical variable $\phi$, $J(\phi)$, imposed  by an auxiliary variable $a$ (Lagrange multiplier),
\begin{eqnarray}
\mathcal{L}'(\phi,a)= \mathcal{L}(\phi,\partial \phi,\partial^2\phi)+a J(\phi) \, , \label{eqn L prime}
\end{eqnarray}
where $\mathcal{L}$ is given by (\ref{eqn L without constraint}) satisfying (\ref{eqn construction L2}), and the obvious choice for $J(\phi)$ that guarantees the low derivative background $\phi_0$ is,
\begin{eqnarray}
J(\phi):=\Theta(\mathcal{L}^{(1)};\phi) \, . \label{eqn source}
\end{eqnarray}
It is clear that $J(\phi)$ is {\it a physical input} as it is  not derived from the unstable theory $\mathcal{L}$. Furthermore, it does not trivialize the dynamics of the variable $\phi$ because the auxiliary ($a$) now becomes dynamical, solving a 2nd-order differential equation, and bearing the higher derivative effects. This will be verified in the next subsection.  

 Equivalently, $J(\phi)$ {\it sources} the auxiliary $a$, and the physical interpretation is clear: the dynamics of the low energy mode (\ref{eqn source}) \textit{sources} any effect requiring higher derivative  terms for its description. Following Dirac's programme in section \ref{sec stability}, we will see the non trivial character of this setup at work with the {\it two physical inputs} (\ref{eqn construction L2}) and (\ref{eqn source}), as it generates more constraints than (\ref{eqn source}) itself. Hence, (\ref{eqn source}) is not identically zero and $\mathcal{L}'$ dynamics can be richer than the corresponding low derivative sector (\ref{eqn def background}). It will reduce the dimensionality of phase space, leaving no ghost, and relating $a$ and $\phi$ such that altogether there is at most one truly dynamical degree of freedom that is in general different than $\phi_0$.

As a first taste of this analysis notice that the constraint (\ref{eqn source}) implies the required degeneracy only on-shell: namely, (\ref{eqn construction L2}) does not vanish identically, but it does vanish when it is valued on solutions to the Euler-Lagrange equation for $a$, 
\begin{eqnarray}
\Theta(\mathcal{L}';a)=J(\phi)=0\, , \label{eqn E-L a}
\end{eqnarray}
which sets $\phi=\phi_0$ (only on-shell) and however, {\it it does not trivialize} $\mathcal{L}'$ because there is an additional dynamical variable ($a$) that is taking care for the observation of these low energy modes amid the higher derivative effects of $\mathcal{L}^{(2)}$. Indeed, $a$ solves a differential equation on its own:

\subsection{A taste of the new  dynamics}\label{sec new d}
The differential equation for the {\it ghost condensate} ($a$),  with the notation (\ref{eqn notation E-L}), is deduced from,
$$\Theta(\mathcal{L}';\phi)= \Theta(\mathcal{L};\phi)+ \Theta\left(a \,  J(\phi);\phi\right)=0\, , $$
where we can use the other Euler-Lagrange equation (\ref{eqn E-L a}),
\begin{eqnarray}
\left.\Theta \left(a \, \Theta(\mathcal{L}^{(1)};\phi);\phi\right)\right\vert_0&=&-\left.\Theta(\mathcal{L}^{(2)};\phi) \right\vert_0 \, . \label{eqn gc}
\end{eqnarray}
More explicit, by assumption (\ref{eqn non degeneracy L1}) the term $\Theta(\mathcal{L}^{(1)};\phi) $ depends linearly on $\ddot{\phi}$ such that, for instance, in the mechanics of a point particle the left hand side of (\ref{eqn gc}) is a second-order differential operator acting on $a$,
$$
\left.\left(\frac{d^2}{dt^2} \frac{\partial}{\partial \ddot{\phi}}-\frac{d}{dt} \frac{\partial}{\partial \dot{\phi}}+\frac{\partial}{\partial \phi}\right)\left(a \, \Theta(\mathcal{L}^{(1)};\phi)\right)\right\vert_0\, ,$$
which may contain damping terms and non vanishing time dependent coefficients valued on the sourcing, low energy mode $\phi_0$, which we denote with the symbol $\vert_0$. 

Finally, the right hand side of (\ref{eqn gc}) is a forcing term driving the  non homogeneous equation for $a$, which is defined by the dynamics of the higher derivative sector $\mathcal{L}^{(2)}$. Notice that it is non-zero because it is valued on $\phi_0$, and also, it is necessarily suppressed by the new-physics scale $\Lambda$ that was introduced with $\mathcal{L}^{(2)}$. 

Prominently, in the case of relativistic field theory, this setup matches the speed of propagation and (acoustic) cone of influence for $a$-wave equation (\ref{eqn gc}) with the wave-equation for the low energy mode (\ref{eqn E-L a}),  even though there are derivative interactions of $\phi_0$ forcing $a$ on the right hand side of (\ref{eqn gc}).

Finally, let us stress: that there are second order equations {\it does not mean} on its own that the system is stable in the sense of Ostrogradsky \cite{tolley}. This requires a critical verification of the energy and dimensionality of phase space \cite{tolley,noui,1woodard}, which we show in section \ref{sec low}.

For more details of the ghost condensate dynamics, see section \ref{sec s condensates}, and in particular section \ref{sec example} for a simple concrete example. There, we solve the ghost condensate $a$ in equation (\ref{eqn sol a}), which exemplifies the prior discussion.

\subsection{A taste of the stability, and the criticality}\label{sec first stability}
Let us review the key aspects that lead to  ``stable'' dynamics  for (\ref{eqn L prime}), while the initial theory (\ref{eqn L without constraint}) is unstable (Here we only refer to ``stability'' only in the sense of Ostrogradsky):
It is easy to see that the conserved quantity derived from the time homogeneity of a second-order time derivative action with Lagrangian $\mathcal{L}$ (which we associate with the energy for a standard low derivative theory), is given by,
\begin{eqnarray}
E_\phi= \frac{\partial \mathcal{L}}{\partial \ddot\phi} \ddot\phi + \frac{\partial \mathcal{L}}{\partial\dot\phi} \dot\phi - \mathcal{L} -   \dot\phi \frac{d}{dt}\frac{\partial \mathcal{L}}{\partial  \ddot\phi}\,. \label{eqn energy E}
\end{eqnarray}
The configuration space is determined by the coordinates $\phi,\dot\phi,\ddot\phi$ and $\dddot\phi$. All terms on the right hand side of (\ref{eqn energy E}) with the exception of the rightmost, depend only on $\phi,\dot\phi$  and $\ddot\phi$  in a non trivial way. However, expanding the total derivative in the rightmost term, it is easy to see that it depends linearly on $\dddot\phi$ as,
\begin{eqnarray}
-\frac{\partial^2 \mathcal{L}}{\partial\ddot\phi^2}\dot\phi\dddot\phi\,.\label{eqn critical term}
\end{eqnarray}
Then, if (\ref{eqn non degeneracy L2}) does not vanish identically and there are no constraints, this linear dependance implies that the energy is not bounded from below \cite{1woodard,2woodard,3woodard,1motohashi,noui,simon, tolley,1langlois,klein,2langlois}. On the other hand, for a higher derivative sector with the structure (\ref{eqn construction L2}), first note that in the energy for the small fluctuations about the low energy mode ($E_\pi$), for the Lagrangian $\mathcal{L}_\pi$ (\ref{eqn quadratic L}) up to order $\mathcal{O}(\epsilon^2)$  this term vanishes,
$$-\left.\frac{\partial^2 \mathcal{L}}{\partial\ddot\phi^2} \right\vert_{0}\dot\pi\dddot\pi\propto -\Theta(\mathcal{L}^{(1)}(\partial \phi_0,\phi_0)) \dot\pi\dddot\pi\,,$$
Second, without restricting to small fluctuations, let us use the full structure of the corrected theory $\mathcal{L}'$: because the energy is expressed in terms of solutions to the equations of motion, we can use again the Euler-Lagrange equation for $a$ (\ref{eqn E-L a}), which imposes the low energy mode that by definition {\it vanishes (\ref{eqn construction L2}) only on-shell, as well as the critical would-be linear term (\ref{eqn critical term}) that is proportional to} $ \dddot{\phi}$ (Note that the linear terms in $a$ in the energy are only temporary, because $a$ is a Lagrange multiplier that must be solved). Similar considerations can be done for a field theory, as we detail in the next sections.

The vanishing of the linear dependance on $\dddot{\phi}$ of the energy amounts to a would-be linear momentum in the Hamiltonian that is constrained to other canonical coordinates; thus, eliminating this strict linear dependance on the hypersurface of constraints. This is explicitly shown in the Hamiltonian analysis with constraints in section \ref{sec stability}. Furthermore, we analyze the reduction of dimensionality of phase space in \ref{sec low}.

Finally, let us note: the energy $E_\phi$ for $\mathcal{L}'$ (\ref{eqn L prime})  also includes the term (\ref{eqn critical term}). Thus, it is easy to see that the condition (\ref{eqn construction L2}) amounts to set the low derivative sector (0-th order solution for the perturbative expansion) as a critical point of the energy,
$$\frac{\partial E_\phi}{\partial \dddot{\phi}}= -\frac{\partial^2 \mathcal{L}'}{\partial\ddot\phi^2}\dot\phi\, \propto \Theta(\mathcal{L}^{(1)};\phi)=0\,. $$ 

\section{Summary of the setup}\label{sec general}

The setup proposed in this letter can be summarized in two observations: first, there is a class of constrained and {\it degenerate} higher derivative theories $\mathcal{L}'\left(\mathcal{L}^{(1)}\right)$ where the Ostrogradsky's instability is removed by the {\it low derivative mode} that is  described with the low derivative sector, $\mathcal{L}^{(1)}$. Second, the low derivative mode does not trivialize the higher derivative effects. Instead, the higher derivative effects become mere corrections to the low derivative dynamics derived from $\mathcal{L}^{(1)} $ (The notation and assumptions on $\mathcal{L}^{(1)} $ were given in section \ref{sec assumptions}).

This class of degenerate higher derivative theories $\mathcal{L}'(\partial^2\phi,\partial\phi,\phi,a)$ satisfies the following two conditions:

\begin{enumerate}
\item{$\mathcal{L}' $ can be written as,
\begin{eqnarray}
\mathcal{L}'(\phi,a)=\mathcal{L}(\phi,\partial\phi,\partial^2\phi) +a\Theta(\mathcal{L}^{(1)};\phi) \, , \label{eqn total lagrangian}
\end{eqnarray}
}
\item{$\mathcal{L}' $ satisfies,
\begin{eqnarray}
\frac{\partial^2 \mathcal{L}'}{\partial \ddot{\phi}^2}=c(\phi) \, \Theta(\mathcal{L}^{(1)};\phi)  \,,  \label{eqn principal clasica}
\end{eqnarray}
where the notation $\Theta(\,\cdot\, ;\phi)$ was defined in (\ref{eqn notation E-L}). $\mathcal{L}^{(1)}(\phi,\partial\phi)$ is a nondegenerate  {\it low derivative} sector in $\mathcal{L} (\phi,\partial\phi,\partial^2\phi) $, and $c(\phi)$ is a non-zero, non-singular  function which may depend on up to second derivatives of $\phi$.
}
\end{enumerate}
In particular, there must be at least one such $\mathcal{L}^{(1)}$-sector that remains after all terms proportional to $ \ddot{\phi}$ and $ a$ are removed from $\mathcal{L}'$. For instance, in the last section we chose a particularly simple case\footnote{Note that for the particularly simple choice (\ref{eqn choice L}) for $\mathcal{L}$, the explicit term $\mathcal{L}^{(1)} $ is redundant with respect to the term $a\Theta(\mathcal{L}^{(1)};\phi) $  to derive the {\it classical} dynamics of $\mathcal{L}'$. Namely, the contribution of the former   to the Euler-Lagrange equation for $\phi$ vanishes because of the Euler-Lagrange equation for $a$. However, the former term does contribute to the energy. Furthermore, for more general $\mathcal{L}$ than (\ref{eqn choice L}), the mentioned redundancy in the classical equations does not necessarily takes place.} in (\ref{eqn total lagrangian}),
\begin{equation}\label{eqn choice L}
\mathcal{L} (\phi,\partial\phi,\partial^2\phi) = \mathcal{L}^{(2)}+ \mathcal{L}^{(1)}\, , 
\end{equation}
where $\mathcal{L}^{(2)}$ is proportional to $\ddot{\phi}$. Further assumptions were given in section \ref{sec assumptions}. A concrete example is given in section \ref{sec example}, in equations (\ref{eqn L1 example}), (\ref{eqn L2 example}) and (\ref{eqn Lprime example}), and at the end of section \ref{sec field} for a real scalar field.\\

The two equations (\ref{eqn total lagrangian}) and (\ref{eqn principal clasica}) are the key ingredients for the class of theories analyzed in this letter. Importantly, as we verify in section \ref{sec stability}, they make the theory $\mathcal{L}'$ degenerate in order to avoid the Ostrogradsky's instability, much in the same way as in general degenerate theories \cite{1langlois}. Note that this construction does not avoid the possibility of  instabilities due to  unbounded potential or other origins different than Ostrogradsky's.

Let us summarize the physical meaning of the setup given by equations (\ref{eqn total lagrangian}) and (\ref{eqn principal clasica}): the term $a\Theta(\mathcal{L}^{(1)};\phi) $ in (\ref{eqn total lagrangian}) imposes
 $$\Theta(\mathcal{L}^{(1)};\phi)=0\, , $$
 {\it only}\footnote{Let us stress that $\Theta(\mathcal{L}^{(1)};\phi) $ nor (\ref{eqn principal clasica}) are identically zero, and only vanish on-shell. Because of this subtle but important fact there are secondary constraints that turn the theory $\mathcal{L}'$ degenerate. See section \ref{sec stability} and equation (\ref{newConstraint}) for more details. In particular, we will resort to the standard symbol of ``weak equality'' ($\approx$). Among other consequences in the dynamics, it is worth to emphasize that the weak equality implies that $\Theta(\mathcal{L}^{(1)};\phi)$ cannot be taken equal to zero in the Lagrangian $\mathcal{L}'$ (\ref{eqn total lagrangian}), neither in the condition (\ref{eqn principal clasica}). Hence, it is possible to integrate the condition (\ref{eqn principal clasica}), whose right hand side is in general non-vanishing, to find a possibly different than quadratic dependance of $\mathcal{L}'$ on $\ddot{\phi}$. We show an example in section \ref{sec example}, equation (\ref{eqn L2 example}).} on-shell. Namely, this term is in charge of fixing the low derivative dynamics of $\mathcal{L}^{(1)}$ as a background at all times, such that there can be an energy regime where this background is more relevant for the dynamics than the deviations caused by the higher derivative effects. That this constraint holds at all times if it holds at an initial time is guaranteed by secondary constraints, and the particular form taken by the Lagrange multipliers, which follow from Dirac's programme to construct the total Hamiltonian. This is explicitly done in the next section.

On the other hand, the specific condition (\ref{eqn principal clasica}) is designed such that altogether with the secondary constraints that were discussed above, the higher derivative sector of the theory $\mathcal{L}'(\phi,a)$ is degenerate. In particular, (\ref{eqn principal clasica})  and these secondary constraints explicitly lead to a  constraint on a would-be linear conjugate momentum (equation  (\ref{newConstraint}) in section \ref{sec stability}), which manifestly eliminates the Ostrogradskian instability.

In this way, the dimensionality of phase space is reduced for the theory $\mathcal{L}'(\phi,a)$, such that there is at most one degree of freedom associated with the single dynamical variable $\phi$, which is not a ghost, and which is in general different from the low derivative degree of freedom propagated with the $\mathcal{L}^{(1)}$ dynamics. We verify all of these assertions in section \ref{sec low}.\\

Now, for the simple case where $\mathcal{L}$ in (\ref{eqn total lagrangian}) takes the form (\ref{eqn choice L}), we also showed that the higher derivative effects are carried by the auxiliary dynamical variable, {\it ghost condensate} ($a$), that solves a second-order equation
\begin{eqnarray}
\left.\Theta \left(a \, \Theta(\mathcal{L}^{(1)};\phi);\phi\right)\right\vert_0=-\left.\Theta(\mathcal{L}^{(2)};\phi) \right\vert_0 \, , \label{eqn gc2}
\end{eqnarray}
sourced by the low derivative mode ($\phi_0$) on the left hand side, which we denote with $\vert_0$. Furthermore, the ghost condensate is also forced by the higher derivative dynamics derived from $\mathcal{L}^{(2)}$ on the right hand side, which is also valued on the low derivative mode, hence the symbol $\vert_0$. The  forcing term in the equation for the ghost condensate is suppressed by a new-physics energy scale that, as we specified in section \ref{sec small fluctuations}, must be introduced with the higher derivative sector $\mathcal{L}^{(2)}$. Hence, the effect of the higher derivative terms in $\mathcal{L}'$ are only  suppressed corrections about the low derivative modes described with $\mathcal{L}^{(1)}$. An explicit simple example is given in section \ref{sec example}.\\

This setup can be generalized to the case of field theory. For instance   if $\phi(x)$ is a real scalar field, the generalization of condition (\ref{eqn principal clasica}) for a relativistic theory is straightforward:
\begin{eqnarray}
\frac{\partial^2 \mathcal{L}'}{\partial\phi_{,\mu\nu} \partial \phi_{,\rho\sigma}}=c^{\mu\nu\rho\sigma}(\phi)  \Theta(\mathcal{L}^{(1)};\phi)\,, \label{eqn principal campos}
\end{eqnarray}
where $\phi_{,\mu\nu}\equiv\partial_\mu\partial_\nu\phi$ and  the  assumptions on $c$ $(\equiv c^{0000})$ are extended to $c^{\mu\nu\rho\sigma}(\phi,\partial\phi,\partial^2\phi)$. An example for a real scalar field is given in \ref{sec field}.

\section{Ostrogradsky's stability, and the fundamentally low derivative dynamics of $\mathcal{L}'$}

In this section we explicitly check that the setup for theories $\mathcal{L}'$ satisfying the two equations (\ref{eqn total lagrangian}), and (\ref{eqn principal clasica}) avoids the Ostrogradsky's instability with the same mechanism as in \cite{1langlois}. We will work in a mechanical context in order to make the analysis clearer, but generalization for a real scalar field is straightforward.

First, we will construct the Hamiltonian for the theories $\mathcal{L}'$ according to Ostrogradsky's formulation \cite{1woodard,2woodard,ostrogradsky}. Because the equations (\ref{eqn total lagrangian}), and (\ref{eqn principal clasica}) defining $\mathcal{L}'$ include constraints on the momenta, there are velocities that cannot be inverted in terms of the conjugate momenta. Thus, the total Hamiltonian must be written following Dirac's programme.

Let us sketch the strategy to verify Ostrogradsky's stability: as it is standard in various works \cite{1woodard,2woodard, tolley,1langlois,1motohashi}, we will recognize the would-be  Ostrogradsky's instability in a conjugate momentum that appears to be linear in the Hamiltonian. However, because of the constraints in $\mathcal{L}'$, there are canonical coordinates that are related. In particular, the would-be linear conjugate momentum at the core of Ostrogradsky's instability is no longer an independent coordinate. In other words, the problematic seemingly linear term is in reality not necessarily linear because there are redundancies.

To show this, we recognize the primary constraint included in $\mathcal{L}'$ (\ref{eqn total lagrangian}) by the term,
$$a\Theta(\mathcal{L}^{(1)};\phi)\, , $$

then, we demand that this primary constraint must hold at all times, which leads to a secondary constraint,

$$\Theta(\mathcal{L}^{(1)};\phi)\approx 0\, ,$$

and the solution of Lagrange multipliers in terms of other canonical coordinates. We have used the standard symbol for {\it weak equality} ``$\approx$'', meaning that the equality ``$=$'' holds only on the hypersurface of all constraints in phase space. In other words, the equality is only taken after computing all related Poisson brackets\footnote{See for instance \cite{tyutin} chapters 1 and 2, and ``Dirac's programme'' for a detailed discussion.}.

Now, at this stage we will note that this secondary constraint also makes of the definition of a conjugate momentum an independent additional constraint (\ref{newConstraint}). This is of course not a coincidence, and takes place only because of the condition (\ref{eqn principal clasica}). Let us stress, that this point of the discussion was reached only because of the first condition on $\mathcal{L'}$, (\ref{eqn total lagrangian}). Thus, let us emphasize, both conditions defining the theories $\mathcal{L'}$, (\ref{eqn total lagrangian}) and (\ref{eqn principal clasica}), are necessary and specially written such that the additional constraint (\ref{newConstraint}) arises in the dynamics by demanding that constraints hold at all times. 

The consequence is that the would-be linear momentum at the core of Ostrogradsky's instability is in fact related to other canonical coordinates on the allowed hypersurface of all constraints in phase space. In other words, the problematic momentum can be expressed on-shell in terms of other canonical coordinates, hence removing the strict linearity in terms of this momentum in the total Hamiltonian, and hence removing the Ostrogradsky's instability. 

Finally, in section \ref{sec low}, we will count the number of independent constraints, the number of propagated degrees of freedom, and we will show that for a class of consistent $\mathcal{L}'$ theories, the dynamics is fundamentally low in derivatives.
Let us stress that this setup does not avoid the possibility of  instabilities due to unbounded potential or other origins different than Ostrogradsky's.

\subsection{Ghost elimination via constraints}\label{sec stability}
 
 The six canonical coordinates corresponding to $a(t)$ and $\phi(t)$ in (\ref{eqn total lagrangian}) are \cite{1woodard, ostrogradsky}:
\begin{eqnarray}
\begin{array}{cccccc}
x_1=\phi &  & p_1&=&\frac{\partial \mathcal{L}'}{\partial \dot{\phi}}&-\frac{d}{dt}\frac{\partial \mathcal{L}'}{\partial \ddot{\phi}} \\
x_2=\dot{\phi} & & p_2&=& \frac{\partial \mathcal{L}'}{\partial \ddot{\phi}}& \\
a=a & & p_a&=& \frac{\partial \mathcal{L}'}{\partial \dot{a}}&
\end{array} \label{eqn 6 canonical coord}
\end{eqnarray}
where the elementary non-zero Poisson brackets are $\{a,p_a\}=1$, $\{x_i,p_j\}=\delta_{ij}$, $i,j=1,2$. The assumption of non-degeneracy (\ref{eqn principal clasica}) leads to a conjugate momentum that depends on the acceleration $\ddot{\phi}$,
\begin{eqnarray}
p_2=p_2(\ddot{\phi},x_2,x_1,a)
\end{eqnarray}
Thus, $\ddot{\phi}$ can be inverted in terms of the canonical coordinates $p_2,\, x_1,\, x_2,\,a$. On the other hand, the $p_1$ conjugate momentum depends linearly on $\dddot{\phi}$, 
\begin{eqnarray}
p_1&=&G(x_1,x_2,a,\dot{a},\ddot{\phi}(p_2,x_1,x_2,a))-\dddot{\phi}\,  \frac{\partial^2 \mathcal{L}'}{\partial \ddot{\phi}^2} \label{eqn p1} \\
p_1&=&G(x_1,x_2,a,\dot{a},\ddot{\phi}(p_2,x_1,x_2,a))- \dddot{\phi}\, c(\phi) \, \Theta(\mathcal{L}^{(1)};\phi) \nonumber
\end{eqnarray}
where we have used the definition of $\mathcal{L}'$ (\ref{eqn principal clasica}) and the explicit form of $G$ is irrelevant for our discussion. Since neither the Lagrangian nor the other canonical coordinates depend on $\dddot{\phi}$, the linear dependance of $p_1$ on $\dddot{\phi}$ will remain linear upon the Legendre transform that gives the Hamiltonian,
\begin{eqnarray}
\mathcal{H}&=&p_1x_2+p_2\ddot{\phi}(p_2,x_1,x_2,a)+\lambda\xi_1 \nonumber \\
&-&\mathcal{L}'(p_2,x_1,x_2,a,\ddot{\phi}(p_2,x_1,x_2,a))\label{eqn hamiltonian 1 stage}
\end{eqnarray}
where $\lambda$ is a Lagrange multiplier for the primary  constraint $\xi_1=p_a$, 
\begin{eqnarray}
\xi_1=p_a\approx 0 \label{eqn primary constraint}
\end{eqnarray}
due to the no dependence on the velocity $\dot{a}$. It is easy to verify that this choice of canonical coordinates and Hamiltonian generates correct (lagrangean) time evolution; hence, (\ref{eqn hamiltonian 1 stage}) is the right functional form for the energy \cite{1woodard,ostrogradsky}.

As has been widely discussed in the literature \cite{1woodard,2woodard,3woodard,noui,simon, tolley,1langlois,klein,1motohashi}, unless a constraint expresses $x_2$ in terms of $p_1$, the term $p_1x_2$ in the Hamiltonian is the most basic signal of the Ostrogradsky's instability. It is linear in the conjugate momentum $p_1$ and it renders the Hamiltonian unbounded from below. Let us see in two steps how the stability arises in this construction:\\

1- {\it Consistency and secondary constraints (Dirac's programme)}: The conservation in time of $\xi_1$ implies the low energy dynamics as a secondary constraint $\xi_2$,
\begin{eqnarray}
\dot{\xi}_1&=&\{\xi_1,\mathcal{H}\}=-\frac{\partial \mathcal{H}}{\partial a}\approx 0 \label{eqn L1 eom as secondary constraint} \\
\dot{\xi}_1&=&-p_2 \frac{\partial \ddot{\phi}}{\partial a}+ \frac{\partial \ddot{\phi}}{\partial a}\frac{\partial \mathcal{L}'}{\partial \ddot{\phi}}+ \Theta(\mathcal{L}^{(1)};\phi) = \Theta(\mathcal{L}^{(1)};\phi)\approx 0\,, \nonumber
\end{eqnarray}
where the equality holds on the hypersurface of constraints. The low energy dynamics $\xi_2= \Theta (\mathcal{L}^{(1)};\phi)$ depends on the canonical coordinates $x_1, \, x_2$ and the acceleration $\ddot{\phi}(p_2,x_1,x_2,a)$ by assumption (\ref{eqn non degeneracy L1}), thus, in principle, on the hypersurface of constraints $\xi_2$ only relates $a$ to the canonical coordinates $x_1$, $x_2$, $p_2$,
\begin{eqnarray}
\xi_2=\xi_2(\ddot{\phi}(p_2,x_1,x_2,a),x_1,x_2)\approx 0 \label{eqn theta form 1} \, .
\end{eqnarray} 
The consistency of the Hamiltonian requires again the time conservation of (\ref{eqn theta form 1}),
 $\dot{\xi}_2\approx 0 \, ,$ 
which does not imply any more constraints by the consistency procedure followed above (It expresses $\lambda=\dot{a}$ in terms of canonical coordinates, because $\xi_2$ depends on $a$, such that $\{\xi_2,\xi_1\}\neq 0$).\\

2- {\it Built-in constraints from the structure of $\mathcal{L}'$}: The distinctive feature of the higher derivative sector (\ref{eqn principal clasica}) at work together with the constrained structure of the theory (\ref{eqn total lagrangian}) is that the secondary constraint (\ref{eqn L1 eom as secondary constraint}) that serves to express the auxiliary variable $a$, and $\dot{a}$ in terms of other canonical coordinates, necessarily  implies a new, independent secondary  constraint $\xi_3$, which is seen in the definition of conjugate momentum (\ref{eqn p1}), 
\begin{equation}
\xi_3=p_1- G=-\dddot{\phi} \, c(\phi) \, \Theta(\mathcal{L}^{(1)};\phi)\approx 0  \, ,\label{newConstraint}
\end{equation}
where $G$ depends on $x_2$ and other canonical coordinates. This is an additional constraint because the sole term containing $\dddot{\phi}$, which required the independent momentum $p_1$  (\ref{eqn p1}) in order to express all time derivatives of $\phi$ in terms of canonical coordinates, vanishes on the hypersurface of constraints as a by-product of $\Theta(\mathcal{L}^{(1)};\phi)\approx 0$. In other words, $p_1$ is no longer independent but related to $x_2,\, p_2,\,x_1$ by means of $G$ (\ref{newConstraint}) on the hypersurface of constraints, which gets rid of the strict linear dependance of $\mathcal{H}$ on $p_1$ (\ref{eqn hamiltonian 1 stage}). In fact, as a new constraint, $\xi_3$ must be added to the set $\{\xi_1,\, \xi_2\}$ to solve suitably all together, and new constraints may appear following Dirac's Programme. For the moment, let us stress that the desired expression of  $x_2$ in terms of $p_1$ is inherent in the structure of $\mathcal{L}'$  and {\it the Ostrogradsky's instability is not present}. Namely, the term that would be linear in an unconstrained higher derivative theory $\sim p_1x_2$, is not necessarily linear for the degenerate, constrained theories $\mathcal{L}'$, 
$$p_1 \, x_2(p_1,\dots)\, .$$
Let us emphasize that the independent constraint $\xi_3$ that relates the would-be linear momentum $p_1$ and other canonical coordinates such as $x_2$ is entirely due to the intimate relation between the structure of the higher derivative sector (\ref{eqn principal clasica}) and the constraint included in (\ref{eqn total lagrangian}) that takes care of  the propagation of the low energy modes. Namely, the term (\ref{eqn source}). 

In other words, had the condition (\ref{eqn principal clasica}) or the constraint imposed by $a$ (low energy mode) (\ref{eqn source}) been different, then  $\xi_3$ would not arise and the Ostrogradsky's instability would still be present. A particular case of such a failure was  shown in \cite{tolley}: namely, consider the case that $a$ imposes a secondary constraint (\ref{eqn source}) that is also a second-order equation for $\phi$, but different than the form $\Theta(\mathcal{L}^{(1)};\phi)$ which is used to define $\mathcal{L}'$ in (\ref{eqn principal clasica}). Then, similarly, there would be a system of two second-order equations of motion, but critically, no additional $\xi_3$ constraint would be present  and Ostrogradsky's instability would remain. The dimensionality of phase space would remain high, signaling the ghost. In short, the constraint (\ref{eqn source}) must be compatible with the structure of $\mathcal{L}'$ (\ref{eqn principal clasica}) such that $\xi_3\approx 0$ emerges. However, as motivated in section \ref{sec small fluctuations}, the special choice (\ref{eqn source}), and (\ref{eqn principal clasica}) leads to the physically  meaningful property that higher derivative effects could be seen as $\Lambda$-suppressed fluctuations about a 0th-order approximation to the full dynamics given by the well known low derivative sector (\ref{eqn source}). An explicit example (\ref{eqn sol a}) is given in section \ref{sec example}.

\subsection{The fundamentally low derivative dynamics of $\mathcal{L}'$, and the amount of degrees of freedom}\label{sec low}

Let us see the reduction of dimensionality in this setup: the special prescription (\ref{eqn principal clasica}) together with the propagation of the low energy mode in (\ref{eqn total lagrangian}) implies in principle a 6-dimensional phase space (\ref{eqn 6 canonical coord}), which is reduced by the 3 independent constraints $\xi_1, \, \xi_2, \, \xi_3$ to at most 3 dimensions. As we saw, it guarantees the elimination of Ostrogradsky's instability, and in accordance with \cite{tolley}, the dimensionality of phase space has been reduced from $4$ in the unstable theory (\ref{eqn L without constraint}) to at most $3$ for $\mathcal{L}'$ defined by (\ref{eqn total lagrangian}), and (\ref{eqn principal clasica}). 

Now, we must continue with Dirac's programme demanding that $\xi_3$ holds at all times,
$$\dot{\xi}_3\approx 0\, .$$ 
As was noted by Dirac for general constrained systems (See for instance \cite{tyutin}), depending on the particular model $\mathcal{L}'$, there are three possibilities at this stage depending on the last equation and the set $\{\xi_1, \, \xi_2, \, \xi_3\}$: 

\begin{itemize}
\item the remaining Lagrange multiplier that multiplies $\xi_3$ in the Hamiltonian is expressed in terms of other canonical coordinates, and there are no more independent constraints,
\item there may arise an inconsistency,
\item new constraints appear on the dynamics $\{\xi_1, \, \xi_2, \, \xi_3,\, \xi_4,\, \dots\}$, and the procedure must be continued. 
\end{itemize}
Let us exclude inconsistent models that directly fit into the second case. On the other hand, for the first case, we would also find an inconsistency because phase space would be 3-dimensional. Then,  let us restrict to non-singular theories with an even-dimensional phase space. The latter can only happen in the last case, where the time conservation of $\xi_3$ implies new constraints. Therefore, we are left with two consistent scenarios: \begin{itemize} 
\item there is more than one additional independent constraint, such that one reaches either an inconsistency at this stage, or the trivial case where there are six independent constraints, and hence, no propagating degree of freedom, $\{\xi_i\}_{i=1, \, \dots \, , \, 6}$ 
\item there is only an additional independent constraint and its time conservation fixes all Lagrange multipliers. Namely, the set of all independent constraints is $\{\xi_i\}_{i=1, \, \dots \, , \, 4}\, .$
\end{itemize}
In the second case, where $\mathcal{L}'$-constrained dynamics does not involve an inconsistency, and there are only 4 constraints phase space is 2-dimensional, which means that there are only two independent initial conditions for the dynamics, and there is only one degree of freedom. This possibility is in accordance with previous analyses \cite{1langlois,1motohashi}.

Furthermore, for the consistent theories $\mathcal{L}'$ where the last case takes place, the dimensionality of phase space and the amount of degrees of freedom would coincide with the low derivative dynamics $\mathcal{L}^{(1)}$. This contrasts to  the unstable, unconstrained theory  (\ref{eqn L without constraint}) that implied a 4-dimensional phase space, or $2$ degrees of freedom, one of them being a ghost. However, let us stress: that the dimensionality of phase space for $\mathcal{L}'$ dynamics reduces to at most the same of the corresponding low derivative theory $\mathcal{L}^{(1)}$ clearly does not imply that the only  propagated degree of freedom is the same as the low derivative mode. Indeed, for $\mathcal{L}'$ dynamics the trajectory in phase space lies on a 2-dimensional hypersurface in a 6-dimensional phase space, while for $\mathcal{L}^{(1)}$ dynamics the trajectory lies in a 2-dimensional phase space.

Finally, to fully understand the physics provided by theories $\mathcal{L}'$, let us restrict to the cases where all the Lagrange multipliers are expressed as functions of other canonical coordinates, and there are only four independent constraints, all of them of second-class. In other words, there was no step where a gauge redundancy had to be fixed, and the Dirac matrix of Poisson brackets between all constraints is regular,
$$\det \Big\vert \{\xi_i,\, \xi_j\} \Big\vert _{i,\, j=\, 1, \, \dots \, , \, 4}\neq 0\, .$$
Then, it is well known\footnote{See for instance Proposition 1 in section 2.3, and Appendix B in \cite{tyutin}.} that it is possible to find equivalent canonical coordinates $(\omega_k,\Omega_l)$, with $k=\, 1, \,2 $ and $l=\, 1, \, \dots \, , \, 4 $, where $\omega_k$ and $\Omega_l$ are separate sets of pairs of canonical coordinates, and only $\omega_k$ are truly dynamical on the trivial 4-dimensional hypersurface of constraints given by $\Omega_l=0$. Furthermore, in these new coordinates, the dynamics can be computed with Poisson brackets with the total Hamiltonian written as,
$$\mathcal{H}(\omega_k,\Omega_l)\vert_{\Omega_l=0}= \mathcal{H}(\omega_k) \, , $$
depending only on one pair of conjugate variables. Hence, there would be only two Hamilton's equations of first-order for $\omega_1,\, \omega_2$. In other words, we see that fundamentally {\it the dynamics for this restricted class of $\mathcal{L}'$ theories does not include higher derivatives} for the sole, truly dynamical degree of freedom.\\

Nevertheless, the prescription of $\mathcal{L}'\left(\mathcal{L}^{(1)}\right)$ defined by (\ref{eqn total lagrangian}), and (\ref{eqn principal clasica}) could still be practical\footnote{Notice, for instance, theories of the type \cite{galileons} which are written in a form with higher derivatives, even though they lead to second order equations of motion.}, because writing the equivalent low derivative theory in terms of $\omega_1,\, \omega_2$ is in general prohibitively difficult. This is specially true for the cases in this letter, because the equation (\ref{eqn principal clasica}) implies that $\mathcal{L}'$ must be nonlinear in $\ddot{\phi}$. We verify this in the next section.

In other words, provided one has a sound grasp describing a physical system in terms of the low derivative degree of freedom $\phi$ with Lagrangian  $\mathcal{L}^{(1)}(\phi)$ in some energy regime, then, it is plausible to explore would-be higher derivative effects of the form $\mathcal{L}'$ (\ref{eqn total lagrangian}), and (\ref{eqn principal clasica}) (which is still fundamentally low in derivatives and ghost-free on top of the $\mathcal{L}^{(1)}$ background), instead of writing an equivalent system in terms of $\omega_1,\, \omega_2$ canonical coordinates, and the corresponding low derivative Lagrangian, which are unknown in the regime where $\phi$ is a good degree of freedom. 

All in all, the setup (\ref{eqn total lagrangian}), and (\ref{eqn principal clasica}) may be more suitable to explore ghost-free effective theories with higher derivative terms associated to the $\phi$ low derivative degree of freedom.
\section{Dynamics of the ghost condensate}\label{sec s condensates}

Below we discuss the generalities of the ghost condensate dynamics that arise in this setup. We stress on the speed of propagation, the (acoustic) cone of influence for the wave equation, as well as  the stability and subluminality properties that can be inherited from a healthy low derivative sector, despite the built-in derivative self-interactions in the higher derivative sector. We start with the simplest case in mechanics of a single particle to show how the low dimensionality of phase space, which is intimately related to the elimination of the ghosty degree of freedom, is indeed tied to the boundedness from below of the energy. We only show a solution of the ghost condensate for this simple case.

\subsection{A first example, the energy, and the dimensionality of phase space}\label{sec example}

The Hamiltonian for the Lagrangian $\mathcal{L}'$ (\ref{eqn total lagrangian}), (\ref{eqn principal clasica}), with the constraint structure in section \ref{sec stability} properly solved, first, would give the correct time evolution for at most one truly dynamical variable (at most a 2-dimensional phase space for a consistent model), and second, it would be bounded from below for the one dynamical variable (no Ostrogradsky's instability). We showed  these two aspects in the last section and how they are intimately linked, in accordance with \cite{tolley}. However, it is difficult to write the Hamiltonian in its explicit form for a particular model because of the built-in self-interactions that are necessary for ghost condensation: indeed, integrating $\mathcal{L}'$ from the condition (\ref{eqn principal clasica}), assuming that $c$ is a polynomial function of $\ddot{\phi}$, and  because $\Theta(\mathcal{L}^{(1)};\phi)$ is linear in $\ddot{\phi}$, there is a highest order for $\ddot{\phi}$ in $\mathcal{L}'$ that scales at least as, 
$$\mathcal{L}^{(2)}\propto \ddot{\phi}^p\, , $$
with $p\geq3$. Hence, although we will not solve all the constraints neither compute the explicit Hamiltonian, we will compute, both the lagrangean dynamics and the energy function, knowing from the Hamiltonian analysis that the low dimensionality of phase space will be intimately tied to the boundedness from below of the energy function. 

Let us consider a first example: take the harmonic oscillator as the low derivative sector,
\begin{eqnarray}
\mathcal{L}^{(1)}=\frac{1}{2}\dot{\phi}^2-m^2\phi^2 \, , \label{eqn L1 example}
\end{eqnarray}
and consider a simple higher derivative sector keeping the symmetry $\phi\rightarrow -\phi$,
\begin{eqnarray}
\mathcal{L}^{(2)}=\frac{1}{2 \Lambda^5}\ddot{\phi}^2\left(-\frac{1}{6}\ddot{\phi}^2+m^4\phi^2 \right) \, , \label{eqn L2 example}
\end{eqnarray}
where $\phi^2,\, \Lambda^{-1},\, m^{-1}$ have units of time, such that in this setup the complete theory takes the form (\ref{eqn total lagrangian}),
\begin{eqnarray}
\mathcal{L}'(\phi,a)=\mathcal{L}^{(2)}+\mathcal{L}^{(1)}+a \Theta(\mathcal{L}^{(1)};\phi) \, , \label{eqn Lprime example}
\end{eqnarray}
where, we have used the notation (\ref{eqn notation E-L}), and the condition (\ref{eqn principal clasica}) holds,
$$\frac{\partial^2 \mathcal{L}'}{\partial \ddot{\phi}^2}= \frac{\partial^2 \mathcal{L}^{(2)}}{\partial \ddot{\phi}^2}= \frac{\left(\ddot{\phi}-m^2\phi\right)}{\Lambda^5}  \Theta(\mathcal{L}^{(1)};\phi) \, . $$
The Euler-Lagrange equation for $a$ and $\phi$ are respectively: the low derivative sector for $\phi(t)$, whose solutions we have denoted as $\phi_0$,
$$\Theta(\mathcal{L}';a)= \Theta(\mathcal{L}^{(1)};\phi)=\left(\frac{d^2}{dt^2}+m^2\right)\phi_0=0 \, , $$
and the differential equation for $a(t)$,
$$\Theta(\mathcal{L}';\phi)= \Theta(\mathcal{L}^{(1)};\phi) + \Theta(\mathcal{L}^{(2)};\phi)+ \Theta\left(a \,  J(\phi);\phi\right) =0 \, ,$$
where the last term with $J(\phi)=\Theta(\mathcal{L}^{(1)};\phi)$ reads,
$$\Theta\left(a \,  J(\phi);\phi\right)=\left(\frac{d^2}{dt^2}\frac{\partial}{\partial \ddot{\phi}}-\frac{d}{dt}\frac{\partial}{\partial \dot{\phi}}+ \frac{\partial}{\partial  \phi}\right)a(-\ddot{\phi}-m^2\phi)\, ,$$
which is a linear differential operator acting on $a(t)$, left hand side of,
\begin{eqnarray}
\left(\frac{d^2}{dt^2}+m^2\right)a=\frac{m^5}{\Lambda^5}\left(3m^3\phi_0^3-4m\phi_0\dot{\phi}_0^2\right) \, , \label{eqn a example} 
\end{eqnarray}
where we have used $\phi=\phi_0$ and the right hand side is the contribution from the higher derivative dynamics $\Theta(\mathcal{L}^{(2)};\phi)\vert_0$ also written in terms of the low derivative solutions, hence the subindex $\vert_0$. As anticipated, the explicit term $\mathcal{L}^{(1)}$ in the total Lagrangian $\mathcal{L}'$ is redundant to derive the {\it classical} equations, however, it does contribute with the standard form of energy, as we verify below.

Note that both equations are of second order. Even though it seems that $4$ initial conditions are required to find particular solutions, $a$ is an auxiliary variable that is linked to the only $2$ initial conditions that fix the particular solution $\phi_0$. Indeed, we  know from the constrained evolution, which is implied by the Hamiltonian, that phase space is at most 2-dimensional. Let us see: the energy function (\ref{eqn energy E}) has a contribution from the harmonic oscillator $\mathcal{L}^{(1)}$, which is the low energy sector,
\begin{eqnarray}
\mathcal{E}^{(1)}=\frac{1}{2}\dot{\phi}_0^2+\frac{m^2}{2}\phi_0^2 \, , 
\end{eqnarray}
and a new contribution from the constrained higher derivative sector,
\begin{eqnarray}
\mathcal{E}^{(2)}=\frac{m^5}{\Lambda^5}\left(2m\phi_0^2\dot{\phi}_0^2+\frac{1}{4}m^3\phi_0^4\right)+\dot{\phi}_0\dot{a}+m^2\phi_0 a \, , \nonumber\\
\end{eqnarray}
where all terms but the last two of the last line (dependent on $a$ and $\dot{a}$) are manifestly positive contributions to the energy $\mathcal{E}= \mathcal{E}^{(1)}+ \mathcal{E}^{(2)}$. Let us recall that $a(t)$ is an auxiliary Lagrange multiplier, and it must be solved: to do so, we showed in section \ref{sec stability} that the  constraint that eliminates the Ostrogradsky's instability (\ref{newConstraint}) is also necessarily associated to the constraint (\ref{eqn L1 eom as secondary constraint}) and derived conditions that allow to express $a$ and $\dot{a}$ in terms of other canonical coordinates. In other words, the particular solution for $a(t)$ (\ref{eqn a example}) is linked to the solution to the other canonical coordinate $\phi=\phi_0$ by the boundedness from below of the energy (elimination of Ostrogradsky's instability (\ref{newConstraint})), such that the dimensionality of phase space is at most $2$. Indeed, if the particular solution for $\phi_0$ is,
$$\phi_0(t)=c_1\, y_1(t)+c_2\, y_2(t)$$
where $c_1, \, c_2$ are fixed by initial conditions and $y_1=\cos(mt)$, $y_2=\sin(mt)$, then, $a(t)$ is written as the solution to the homogeneous equation of (\ref{eqn a example}), namely, with the same basis of independent functions $\{y_1,y_2\}$, plus a particular solution to the non homogeneous equation (\ref{eqn a example}) carrying the higher derivative effects from $\mathcal{L}^{(2)}$, $N(t)$,
$$a(t)= c_3\, y_1(t)+c_4\, y_2(t) +\frac{1}{\Lambda^5}N(t)\, ,$$
where,
$$\frac{1}{\Lambda^5}N(t)=-\frac{1}{m}y_1\int dt \,  y_2 \, \left.\Theta(\mathcal{L}^{(2)};\phi)\right\vert_0+ (y_1 \leftrightarrow y_2)\, ,$$
depends only on $c_1$ and $c_2$, and $\left.\Theta(\mathcal{L}^{(2)};\phi)\right\vert_0 $ is given by the right hand side of (\ref{eqn a example}). The key aspect is: we can find $c_3$ and $c_4$ in terms of $c_1$ and $c_2$ just by writing the energy, which must be free of linear instabilities (bounded from below), {\it as the Hamiltonian analysis revealed} in section \ref{sec stability}. Let us see: the last two terms in $\mathcal{E}^{(2)}$ are expressed with these solutions as,
$$\dot{\phi}_0\dot{a}+m^2\phi_0 a=m^2\left(c_1\, c_3+c_2 \, c_4\right) (y_1^2+y_2^2) + \frac{\dot{\phi}_0\dot{N}+m^2\phi_0 N}{\Lambda^5} \, , $$ 
where the quadratic first term must be positive definite for every particular solution of $\phi_0(t)$. This defines $\{c_3,\, c_4\}$, and hence $a(t)$ in terms of $\{c_1,\, c_2\}$ in the lagrangean dynamics. For instance, $c_3\equiv c_1$, and $c_4 \equiv c_2$. That there is such a form of the energy linking the boundedness from below to the low dimensionality of phase space (namely, solutions depending only on $\{c_1,c_2\}$) is a signature of the elimination of Ostrogradsky's instability, which would be {\it manifest in the constrained Hamiltonian} time evolution. The last two terms dependent on $N(t)$, and $\dot{N}(t)$ are non linear functions of $\, y_1,\, y_2$ and integrals, and thus, are not critical in this analysis (however, they could show other types of instabilities depending on the particular model).\\

All in all, the ghost condensate ($a$) takes the form of a $\Lambda$-suppressed correction ($N(t)$) superposed to the low energy mode $\phi_0$ (up to a linear combination of $\{y_1,y_2\}$ also conditioned to the boundedness from below of the energy, as was  analyzed above),
\begin{eqnarray}
a= \phi_0+\frac{1}{\Lambda^5}N \,, \label{eqn sol a}
\end{eqnarray}
such that the ghost condensate carrying  the $\Lambda^{-5}N(t)$ higher derivative effects is also specified by the preparation of  the low energy mode ($\phi_0$) with {\it two} initial conditions.  This is reminiscent of the Ansatz ($\phi=\phi_0+\epsilon \pi$) for the perturbative expansion about $\phi_0$ in $\mathcal{L}'$ dynamics, which motivated this setup in section \ref{sec small fluctuations}, where we identify the solution to all orders
\begin{eqnarray}
\epsilon\pi=\Lambda^{-5}N\, .\label{eqn pi}
\end{eqnarray}
Let us stress that this form of the solution for the ghost condensate  is not restricted to the particular higher derivative sector (\ref{eqn L2 example}). In fact, the form of the ghost condensate as a superposition of a $\Lambda$-suppressed fluctuation atop the low energy mode (\ref{eqn sol a}) holds for any higher derivative sector $\mathcal{L}^{(2)}$ compatible in the form (\ref{eqn principal clasica}) to a quadratic low derivative sector $\mathcal{L}^{(1)}(\dot{\phi}^2,\phi^2)$. This follows in such a simple way because the equations for $a(t)$ and $\phi(t)$ share the same {\it linear}  differential operator, $(d^2/dt^2+m^2)$. More intricate relations arise with nonlinearities in the low derivative sector $\mathcal{L}^{(1)}$.

\subsection{Scalar ghost condensate, the wavefront velocity  
and the causal structure}\label{sec field}
Ghost condensates in this setup inherit many of the features of propagation of the low energy modes. This holds because the differential operator for the ghost condensate is derived from the dynamics of the low energy mode ($\Theta(\mathcal{L}^{(1)};\phi)$), and most prominently, because the effective metric coincides for both field  equations. We can easily see this: consider the complete theory with the form (\ref{eqn total lagrangian}),
$$\mathcal{L}'(\phi,a)=\mathcal{L}^{(2)}+\mathcal{L}^{(1)}+a \Theta(\mathcal{L}^{(1)};\phi) \, , $$
where  the low derivative sector is the most general self-interacting real scalar field $\phi(t,\vec{x})$ with Lorentz invariant Lagrangian $\mathcal{L}^{(1)}$ that depends only on powers of $\partial_\mu\phi\partial^{\mu}\phi$ and $\phi$. The equation of motion for $\phi$ is derived from,
$$\Theta(\mathcal{L}';a)=\Theta(\mathcal{L}^{(1)};\phi)=0 \, ,$$ and defining $2\mathcal{X}= \partial_\mu\phi\partial^{\mu}\phi $, with $g^{\mu\nu}$ as flat space-time metric, it can be written as,
\begin{eqnarray}
-G^{\mu\nu}\partial_\mu\partial_\nu\phi-2\mathcal{X}\frac{\partial^2 \mathcal{L}^{(1)}}{\partial \phi \partial \mathcal{X}}+\frac{\partial \mathcal{L} ^{(1)}}{\partial \phi}=0 \, ,  \label{eqn L1 eom}
\end{eqnarray}
where $G^{\mu\nu}$ depends on $\phi$ and its first derivatives,
\begin{eqnarray}
G^{\mu\nu}=\frac{\partial \mathcal{L}^{(1)}}{\partial \mathcal{X}}g^{\mu\nu}+ \frac{\partial^2 \mathcal{L} ^{(1)}}{\partial \mathcal{X}^2}\partial^\mu\phi\partial^\nu\phi \,. \label{eqn effective metric}
\end{eqnarray}
$G^{\mu \nu}$ defines the characteristic curves of the field equation and the propagating character of solutions to (\ref{eqn L1 eom}). Namely, whether the equation is hyperbolic, parabolic or elliptic. In the case it is hyperbolic, there are indeed propagating solutions. In other words, the characteristic curves  are real and they serve to identify the  wavefront and its velocity \cite{shore}. In short, $G^{\mu\nu}$ fixes the speed of sound for the wave equation (\ref{eqn L1 eom}) and the acoustic cone of influence \cite{shore,mukhanov,visser}. Hence, it usually receives the name of effective, or emergent metric. 

On the other hand, the equation for the scalar ghost condensate ($a(x)$) is derived from $\Theta(\mathcal{L}';\phi)=0$. Denoting with $\phi_0$ the solutions to (\ref{eqn L1 eom}), the equation for $a(x)$ takes the form,
\begin{eqnarray}
\left.\Theta \left(a \, \Theta(\mathcal{L}^{(1)};\phi);\phi\right)\right\vert_0&=&-\left.\Theta(\mathcal{L}^{(2)};\phi) \right\vert_0 \, . \label{eqn gc 2}
\end{eqnarray}
The left hand side is a second order differential operator acting on $a(t,\vec{x})$,
$$\left.\left(\partial_\mu\partial_\nu\frac{\partial}{\partial \left(\partial_\mu\partial_\nu\phi\right)}-\partial_\mu\frac{\partial}{\partial \left(\partial_\mu\phi\right)}+\frac{\partial}{\partial \phi}\right)a \, \Theta(\mathcal{L}^{(1)};\phi)\right\vert_0\, ;$$
as $\Theta(\mathcal{L}^{(1)};\phi)$ (left hand side of (\ref{eqn L1 eom})) depends linearly on $\partial_\mu\partial_\nu \phi$, it takes the form,
\begin{eqnarray}
\left(G^{\mu\nu}\partial_\mu\partial_\nu +v^{\mu}\partial_\mu +M^2 \right)a\, , \label{eqn L2 eom}
\end{eqnarray}
where we encounter the same effective metric $G^{\mu\nu}(\phi_0)$. Thus we can identify the same characteristic curves for the field equation of the ghost condensate as for the respective low derivative theory (\ref{eqn L1 eom}). In other words, the higher derivative effects in this setup do not modify the speed of propagation, neither the (acoustic) cone of influence of the low derivative theory, because necessarily, the principal part of the differential operator ($G^{\mu\nu}\partial_\mu\partial_\nu$) is kept invariant by the built-in constraint (\ref{eqn source}). Although this built-in property is what keeps the second-order equations of motion in this setup, thus helping to eliminate the ghost, these features do not arise from the critical definition of the higher derivative sector (\ref{eqn principal clasica}) that is necessary to keep the low dimensionality of phase space and fully eliminate the ghost, as we verified in the previous example. Therefore, although the coincidence of the causal structures follow from the necessary constraint (\ref{eqn source}), this property can be interpreted only in part as a by-product of the setup. All in all, if the effective metric $G^{\mu\nu}$ implied by the low derivative sector $\mathcal{L}^{(1)}$ of the corresponding higher derivative theory $\mathcal{L}'$ satisfies the hyperbolicity, stability and subluminality conditions that were recognized long ago by Aharonov, Komar and Susskind \cite{susskind} (Appendix A),
\begin{eqnarray}
\begin{array}{ccc}
\frac{\partial \mathcal{L}^{(1)}}{\partial \mathcal{X}}>0 \, , & \frac{\partial^2 \mathcal{L}^{(1)}}{\partial \mathcal{X}^2}\geq0 \,, & \frac{\partial \mathcal{L}^{(1)}}{\partial \mathcal{X}} +2\mathcal{X} \frac{\partial^2 \mathcal{L}^{(1)}}{\partial \mathcal{X}^2}>0 \, ,
\end{array}\nonumber
\end{eqnarray}
then, the scalar ghost condensate inherits these properties. The higher derivative sector $\mathcal{L}^{(2)}$ with the structure (\ref{eqn principal campos}), constrained by (\ref{eqn source}) is limited to force the condensate $a(x)$ as in the right hand side of (\ref{eqn gc 2}), but not to define the propagating character of solutions.

On the other hand, considering for definiteness the field theory in four dimensions, the mass term for the ghost condensate is,
$$M^2(\phi_0)=-\left.\Theta \left(\Theta(\mathcal{L}^{(1)};\phi);\phi\right)\right\vert_0\, ,$$
and the damping term is,
$$v^\mu(\phi_0)=\left.\left(2\partial_\nu G^{\mu\nu}+\frac{\partial \Theta(\mathcal{L}^{(1)};\phi)}{\partial \left(\partial_\mu \phi\right)}\right)\right\vert_0 \, ,$$
such that $v^{\mu}$ vanishes if the low derivative sector $\mathcal{L}^{(1)}$ contains no derivative self-interactions. 
Consider an analogous example to (\ref{eqn L1 example}) and (\ref{eqn L2 example}), where the low derivative sector $\mathcal{L}^{(1)}$ is the massive real scalar field ($\phi$) and,
$$\mathcal{L}^{(2)}=\frac{1}{2\Lambda^8} (\Box\phi)^2\left(-\frac{(\Box\phi)^2}{6}+m^4\phi^2\right)\,,$$
satisfies (\ref{eqn principal campos}). $\phi,\,\Lambda$ and $m$ have dimension of mass. With $\mathcal{L}'$ in the standard form (\ref{eqn total lagrangian}), the equation for the scalar ghost condensate $a(x)$ is,
$$\Box{a}+m^2a =\frac{m^5}{\Lambda^8}(3m^3\phi_0^{3}-4m\,\phi_0\,\partial_{\mu}\phi_0\partial^\mu\phi_0)\,,$$
where $\phi_0$ are solutions to the Klein-Gordon equation. Let us stress that the metric is flat for the $a(x)$ field equation, and the speed of light is not endangered by the derivative self-interactions of the low energy mode ($\phi_0$) on the right hand side.

 This contrasts to the typical (unconstrained) low derivative self-interactions that can be obtained for a real scalar with Lagrangian $\mathcal{L}^{(1)}$, whose non-perturbative effects can have disastrous consequences such as superluminality \cite{2arkani-hamed,mukhanov,susskind,shore,visser,2derham,ellis}.  
This holds in general: namely, in the case that $\mathcal{L}^{(1)}$ contains no derivative self-interactions, the right hand side of (\ref{eqn gc 2}) still contains derivative interactions of the low energy mode forcing the ghost condensate, which are induced by the high derivative sector $\mathcal{L}^{(2)}$, and however, do not enclose contributions to the metric  that could potentially spoil causality, or generate other undesirable effects (See related discussions in \cite{2arkani-hamed,mukhanov,susskind,shore,visser,2derham,ellis}).

\section{Discussion: an interpretation of the setup}\label{sec discussion}

The strategy of eliminating ghosts using degeneracies poses a question: how to interpret the {\it ad hoc} constraint and the auxiliary variable?

In this letter, we considered a restricted class of models where such a question has an unified answer. Namely, in section \ref{sec class}, we identified the higher derivative terms that can be made degenerate via the constraint ($J(\phi)$) (\ref{eqn source}) (and its time conservation), which holds the solutions to the low derivative theory as a relevant contribution to the dynamics. 

Furthermore, $J(\phi)$ was interpreted as a source for the higher derivative effects, which are embedded on-shell in the {\it auxiliary variable} $a$, ghost condensate. Indeed, the constraint term in the Lagrangian (\ref{eqn total lagrangian}) takes the form,
$$a\,J(\phi)\, .$$
In other words, the dynamics of the low energy mode ($J(\phi)$) sources any effect requiring higher derivative terms for its description, here embedded in the ghost condensate $a$. Thus, briefly, the answer is: in this setup, the constraint is a source for the new higher derivative effects that are embedded in a low derivative auxiliary variable ($a$).

Finally, the technical question of ``why such a constraint must be imposed?'' is pushed to the fundamental fact that we observe degrees of freedom that follow the second order equation,
\begin{eqnarray}
J(\phi)=0\, , \label{eqn source0}
\end{eqnarray}
at least at low energies, and conceivably, this also holds even if new effects are probed in other energy regimes with subleading contributions from higher derivative terms. 

In other words, the observation of stable low derivative degrees of freedom, relevant for low energy observers, amid the likely unstable effects of  higher derivative terms can be treated as a built-in constraint ($J(\phi)$) in the dynamics. This constraint effectively turns degenerate a class of higher-derivative sectors and does not  trivialize their effects. 

Here, ``nontrivial'' is understood as follows: the ghost condensate ($a$) obeys a novel second order equation (\ref{eqn gc2}) with variable coefficients fixed by the background low derivative modes that solve (\ref{eqn source0}), and it is also forced by the novel higher derivative terms prescribed by   (\ref{eqn principal clasica}).

\section{Conclusions}\label{sec conclusions}

We showed a new class of interaction terms with higher derivatives that can be added to every low derivative real scalar, such that the theory for the higher derivative variable is degenerate, and the nonlinear equation of motion remains of second order. In other words, we give a restriction on the type of higher derivative terms that do not carry ghosts even at non-perturbative level, once a physically motivated constraint is imposed. Hence, they may be more suitable to explore higher order terms in effective theories.

We stressed on the physical significance of the setup. In particular, the necessary constraints that turn the theory degenerate and eliminate ghosts also have a clear physical motivation: they impose and preserve at all times the low derivative dynamics of the higher derivative variable as a background, on top of which, the fluctuations induced by these higher derivative terms are degenerate, and ghost-free. The phenomenological motivation to do so is that there can be an energy regime where the background -low derivative sector- is more relevant for the dynamics than the ghost-free fluctuations caused by the higher derivative terms; in brief, to account for the phenomenological success of low derivative theories at least at low energies. 

It was also shown that this class of constrained theories are fundamentally equivalent to a  conventional low derivative theory, which can be explicitly seen  if the theory is written in terms of a different set of canonical coordinates. However, it was argued that this different set of coordinates may have two disadvantages: first, the equivalent low derivative theory may take a  prohibitively difficult form. Second, this set of coordinates may lack obvious physical significance at the energy regime where the initial low derivative degrees of freedom are useful to describe the physical system.

We showed specific examples where the modifications of the higher derivative effects are suppressed by the new-physics energy scale, even at non-perturbative level. Thus, when the new scale is arbitrarily large, we recovered the initial low derivative theory.

Finally, we specialized to real scalar fields and stressed on the possibly benign properties for the propagating character of solutions for these necessarily {\it constrained} second order derivative self-interactions, as opposed to the standard, {\it unconstrained}, first order derivative self-interactions. In particular, for the former, constrained case, there is no issue with modifications of the effective metric with respect to the low derivative background. We emphasized on the (acoustic) cone of influence for the wave equation, and subluminality despite the built-in derivative self-interactions.\\

\section*{Acknowledgements}
JLS thanks {\it Fundaci\'on ONCE} for support with the grant {\it Programa de Apoyo al Talento} 2019-2020. MVV thanks for the  support from MCQST-DFG.


\begin{thebibliography}{10}

\bibitem{simon}
J.~Z. Simon, ``Higher-derivative lagrangians, nonlocality, problems, and
  solutions,'' {\em Physical Review D}, vol.~41, no.~12, p.~3720, 1990.

\bibitem{1woodard}
R.~P. Woodard, ``The theorem of ostrogradsky,'' {\em arXiv preprint
  arXiv:1506.02210}, 2015.

\bibitem{2woodard}
R.~Woodard, ``Avoiding dark energy with 1/r modifications of gravity,''
  pp.~403--433, 2007.

\bibitem{3woodard}
D.~Eliezer and R.~Woodard, ``The problem of nonlocality in string theory,''
  {\em Nuclear Physics B}, vol.~325, no.~2, pp.~389--469, 1989.

\bibitem{noui}
A.~Ganz and K.~Noui, ``Reconsidering the ostrogradsky theorem:
  Higher-derivatives lagrangians, ghosts and degeneracy,'' {\em arXiv preprint
  arXiv:2007.01063}, 2020.

\bibitem{tolley}
T.-j. Chen, M.~Fasiello, E.~A. Lim, and A.~J. Tolley, ``Higher derivative
  theories with constraints: exorcising ostrogradski's ghost,'' {\em Journal of
  Cosmology and Astroparticle Physics}, vol.~2013, no.~02, p.~042, 2013.

\bibitem{molina}
X.~Ja{\'e}n, J.~Llosa, and A.~Molina, ``A reduction of order two for
  infinite-order lagrangians,'' {\em Physical Review D}, vol.~34, no.~8,
  p.~2302, 1986.

\bibitem{trodden}
A.~R. Solomon and M.~Trodden, ``Higher-derivative operators and effective field
  theory for general scalar-tensor theories,'' {\em Journal of Cosmology and
  Astroparticle Physics}, vol.~2018, no.~02, p.~031, 2018.

\bibitem{taiwan}
T.-C. Cheng, P.-M. Ho, and M.-C. Yeh, ``Perturbative approach to higher
  derivative and nonlocal theories,'' {\em Nuclear Physics B}, vol.~625,
  no.~1-2, pp.~151--165, 2002.

\bibitem{glavan}
D.~Glavan, ``Perturbative reduction of derivative order in eft,'' {\em Journal
  of High Energy Physics}, vol.~2018, no.~2, p.~136, 2018.

\bibitem{carsten}
C.~G. Knetter, ``Effective lagrangians with higher derivatives and equations of
  motion,'' {\em Physical Review D}, vol.~49, no.~12, p.~6709, 1994.

\bibitem{burgess}
C.~Burgess and M.~Williams, ``Who you gonna call? runaway ghosts, higher
  derivatives and time-dependence in efts,'' {\em Journal of High Energy
  Physics}, vol.~2014, no.~8, p.~74, 2014.

\bibitem{galileons}
A.~Nicolis, R.~Rattazzi, and E.~Trincherini, ``Galileon as a local modification
  of gravity,'' {\em Physical Review D}, vol.~79, no.~6, p.~064036, 2009.

\bibitem{1langlois}
H.~Motohashi, K.~Noui, T.~Suyama, M.~Yamaguchi, and D.~Langlois, ``Healthy
  degenerate theories with higher derivatives,'' {\em Journal of Cosmology and
  Astroparticle Physics}, vol.~2016, no.~07, p.~033, 2016.

\bibitem{klein}
R.~Klein and D.~Roest, ``Exorcising the ostrogradsky ghost in coupled
  systems,'' {\em Journal of High Energy Physics}, vol.~2016, no.~7, p.~130,
  2016.

\bibitem{1derham}
C.~De~Rham and A.~Matas, ``Ostrogradsky in theories with multiple fields,''
  {\em Journal of Cosmology and Astroparticle Physics}, vol.~2016, no.~06,
  p.~041, 2016.

\bibitem{2langlois}
D.~Langlois and K.~Noui, ``Degenerate higher derivative theories beyond
  horndeski: evading the ostrogradski instability,'' {\em Journal of Cosmology
  and Astroparticle Physics}, vol.~2016, no.~02, p.~034, 2016.

\bibitem{lovelock}
D.~Lovelock, ``The einstein tensor and its generalizations,'' {\em Journal of
  Mathematical Physics}, vol.~12, no.~3, pp.~498--501, 1971.

\bibitem{horndeski}
G.~W. Horndeski, ``Second-order scalar-tensor field equations in a
  four-dimensional space,'' {\em International Journal of Theoretical Physics},
  vol.~10, no.~6, pp.~363--384, 1974.

\bibitem{deffayet}
C.~Deffayet, X.~Gao, D.~A. Steer, and G.~Zahariade, ``From k-essence to
  generalized galileons,'' {\em Physical Review D}, vol.~84, no.~6, p.~064039,
  2011.

\bibitem{3langlois}
J.~Gleyzes, D.~Langlois, F.~Piazza, and F.~Vernizzi, ``New Class of Consistent Scalar-Tensor Theories,'' {\em Physical Review Letters}, vol.~114, no.~21, p.~211101,
  2015.

\bibitem{ostrogradsky}
M.~Ostrogradsky, Mem. Ac. St. Petersburg, vol.~6, p.~385, 1850.

\bibitem{1motohashi}
H.~Motohashi and T.~Suyama, ``Third order equations of motion and the
  ostrogradsky instability,'' {\em Physical Review D}, vol.~91, no.~8,
  p.~085009, 2015.

\bibitem{susskind}
Y.~Aharonov, A.~Komar, and L.~Susskind, ``Superluminal behavior, causality, and
  instability,'' {\em Physical Review}, vol.~182, no.~5, p.~1400, 1969.

\bibitem{mukhanov}
E.~Babichev, V.~Mukhanov, and A.~Vikman, ``k-essence, superluminal propagation,
  causality and emergent geometry,'' {\em Journal of High Energy Physics},
  vol.~2008, no.~02, p.~101, 2008.

\bibitem{shore}
G.~Shore, ``Superluminality and UV completion,'' {\em Nuclear Physics B},
  vol.~778, no.~3, pp.~219--258, 2007.

\bibitem{visser}
C.~Barcel{\'o}, S.~Liberati, and M.~Visser, ``Analogue gravity,'' {\em Living
  reviews in relativity}, vol.~14, no.~1, p.~3, 2011.

\bibitem{2arkani-hamed}
A.~Adams, N.~Arkani-Hamed, S.~Dubovsky, A.~Nicolis, and R.~Rattazzi,
  ``Causality, analyticity and an IR obstruction to UV completion,'' {\em
  Journal of High Energy Physics}, vol.~2006, no.~10, p.~014, 2006.

\bibitem{2derham}
C.~de~Rham and H.~Motohashi, ``Caustics for spherical waves,'' {\em Physical
  Review D}, vol.~95, no.~6, p.~064008, 2017.

\bibitem{ellis}
G.~F. Ellis, R.~Maartens, and M.~A. MacCallum, ``Causality and the speed of
  sound,'' {\em General Relativity and Gravitation}, vol.~39, no.~10,
  pp.~1651--1660, 2007.

\bibitem{1arkani-hamed}
N.~Arkani-Hamed, H.-C. Cheng, M.~A. Luty, and S.~Mukohyama, ``Ghost
  condensation and a consistent infrared modification of gravity,'' {\em
  Journal of High Energy Physics}, vol.~2004, no.~05, p.~074, 2004.

\bibitem{tyutin}
G.~Dmitri, and I.~Tyutin, ``Quantization of fields with constraints,'' {\em
  Springer Science \& Business Media}, vol.~2012.


\end{thebibliography}

\end{document}